\documentclass{article}
\pdfoutput=1
\usepackage{jcappub}
\usepackage{amsmath, amssymb, graphics, epsfig}
\usepackage{subfig}
\usepackage{sidecap}
\usepackage{graphicx}
\usepackage{enumerate}
\usepackage[colorinlistoftodos]{todonotes}
\usepackage{color}
\usepackage{xcolor}
\usepackage{multirow}
\usepackage{caption}
\usepackage{capt-of}
\usepackage{float}
\usepackage{url}
\usepackage{comment}
\newcommand{\mathsym}[1]{{}}

\usepackage{cancel}
\usepackage{soul}

\usepackage{xspace}

\newcommand{\hmpc}{$h^{-1}\mathrm{Mpc}$\xspace}
\newcommand{\lcdm}{$\Lambda$CDM\xspace}

\newcommand{\ie}{\textit{i.e.}\xspace}
\newcommand{\eg}{\textit{e.g.}\xspace}

\title{The Void Size Function in Dynamical Dark Energy Cosmologies}

\author[a,b,c]{Giovanni Verza,}
\author[d]{Alice Pisani,}
\author[e,c,f]{Carmelita Carbone,}
\author[g]{Nico Hamaus}
\author[c,f,h,i]{and Luigi Guzzo}
\affiliation[a]{Dipartimento di Fisica e Astronomia ``G. Galilei'', Universit\`{a} degli Studi di Padova, via Marzolo 8,
I-35131, Padova, Italy}
\affiliation[b]{INFN, Sezione di Padova, via Marzolo 8, I-35131, Padova, Italy}
\affiliation[c]{Dipartimento di Fisica ``Aldo Pontremoli'', Universit\`{a} degli Studi di Milano, via Celoria 16, I-20133 Milano, Italy}
\affiliation[d]{Princeton University, Department of Astrophysical Sciences, 4 Ivy Lane Princeton, NJ, 08544, USA}
\affiliation[e]{INAF-IASF, via Corti 12, I-20133 Milano, Italy}
\affiliation[f]{INFN, Sezione di Milano, via Celoria 16, I-20133 Milano, Italy}
\affiliation[g]{Universit\"{a}ts-Sternwarte M\"{u}nchen, Fakult\"{a}t f\"{u}r Physik,
Ludwig-Maximilians Universit\"{a}t, Scheinerstr. 1, 81679 M\"{u}nchen, Germany}
\affiliation[h]{INAF --- Osservatorio Astronomico di Brera, via Brera 28, I-20122 Milano}
\affiliation[i]{INAF --- Osservatorio Astronomico di Brera, via E. Bianchi 46, I-23807 Merate, Italy}

\emailAdd{giovanni.verza@pd.infn.it}
\emailAdd{apisani@astro.princeton.edu}
\emailAdd{carmelita.carbone@inaf.it}
\emailAdd{n.hamaus@physik.lmu.de}
\emailAdd{luigi.guzzo@unimi.it}

\abstract{
We test a theoretical description of the void size distribution function against direct estimates from halo catalogues of the DEMNUni suite of large cosmological simulations. Besides standard \lcdm, we consider deviations of the dark energy equation of state from $w=-1$, corresponding to four combinations in the popular Chevallier-Polarski-Linder parametrisation: $w_0=-0.9;-1.1$, $w_a=-0.3
;0.3$. The theoretical void size function model, relying on the Sheth \& van de Weygaert double barrier excursion set formalism, provides an accurate description of the simulation measurements for the different dark energy models considered, within the statistical errors.
The model remains accurate for any value of the threshold for void formation $\delta_\mathrm{v}$. Its robust consistency with simulations demonstrates that the theoretical void size function can be applied to real data as a sensitive tool to constrain dark energy.
}
\begin{document}

\maketitle

\section{Introduction}
Over the last two decades a revolutionary picture of our Universe emerged: the Hubble diagram of Type Ia supernovae \citep{supernovae1998,perlmutter_supernovae} indicates a recent re-acceleration of the expansion rate, confirmed by other cosmological measurements \cite[\eg][]{zhao_eBOSS2015,zhai_eBOSS2016}. This suggests the existence of an extra fluid, dubbed ``dark energy'', contributing by about 68\% to the total energy density of the Universe, in the simplest model represented by Einstein's cosmological constant $\Lambda$. While the nature of this fluid (and more in general of the dark ingredients dominating the standard cosmological model) remains mysterious, different, complementary probes promise to unveil its properties. Various statistics of the large-scale structure (LSS) of the Universe are key probes upon which we place our hopes to explain the ingredients of the standard model. Galaxy clustering, in particular, provides a way to measure both the expansion history $H(z)$, via baryonic acoustic oscillations (BAO) \citep{6dF_BAO2011,boss2013}, and the growth rate of structure, through the anisotropy of clustering in redshift space \citep[\eg][]{kaiser1987,peacock2001,guzzo2008}.

In this work we focus on a different feature of LSS known as {\sl cosmic voids}. These underdense regions represent the largest structures evident in 3D galaxy maps, with sizes spanning from tens to hundreds of Mpc. Dark energy dominates the mass-energy budget inside voids earlier than for any other structure in the Universe \citep{Pisani2019}. This means that voids reproduce the low-matter density condition characterising the $\Lambda$-dominated era of accelerated expansion before other regions of the cosmic web, making them particularly sensitive to the effects of dark energy \citep{martel1990,lee_DE_2009,bos2012,pisani2015}. With the increasing size of galaxy redshift surveys in recent years, voids are emerging as an effective new probe of cosmology \citep{goldberg2004,park_lee2007,colberg_2008,biswas2010,pisani2015,VIDE,hamaus_constraints_voids,sutter2013aVoid_galaxy,pollina2017,Pisani2019}.

Cosmic voids offer also a laboratory to study physics beyond the standard model \citep{peebles2001,peebles_nusser2010}: their properties, for instance, are sensitive to modified gravity and the sum of neutrinos masses \citep{odrzywolek2009,li2011MG,li-efstathiou2012,clampitt2013,spolyar2013,carlesi2014,zivick_2015_grav,barreira_grav2015,massara_2015,pollina2016,voivodic_2016,cautun_2017,kreisch_neutrini2018,schuster2019_neutrino_voids,perico_2019}. With sizes spanning a large range of values, void statistics respond to different physical mechanisms on different scales. Voids can be affected by environmental influences, such as the large scale tidal force field, playing a dominant role in the evolution of the void’s structure, or the large scale correlations and alignments (see \citep{platen2008,vandeWeygaert2016} and references therein). They have been shown to bear the signature of the Alcock-Paczy{\'n}ski effect \citep{alcock_paczynski,lavaux2012_APtest,sutter2012_APtest,sutter2014d_APtest,hamaus_2014_APtest,hamaus_constraints_voids,mao_AP_sdss}, redshift space distortions (RSD) \citep{martel1990,dekel1994,ryden_voids_rsd,paz_rsd_2013,pisani2014_rsd,pisani2015_rds,hamaus_rsd_2015,cai_rsd2016,achitouv_2017,hamaus_rsd_2017,hawken_vipers_voids}, BAO \citep{kitaura_voids_BAO2017}, CMB lensing and weak gravitational lensing \citep{thompson1987,granett2008,melchior2014,cai2014,calmpit_jain,sanchez2017,cai2017,dong2018_lens,fang2019}, and the integrated Sachs-Wolfe effect \citep{granett2008,ilic2013,kovacs2015,kovacs_garcia,nadathur_isw2016,naidoo2017,kovacs2017,kovacs2019}. 
Void statistics also provide access to higher-order clustering information, providing in this way measurements that are complementary to 2-point statistics from galaxy surveys and CMB anisotropies. As such, voids have the potential to help lift existing degeneracies between different cosmological parameters. Cosmic voids are also relevant for astrophysical studies, offering a unique framework to study the relationship between galaxies and their environment \citep{vogeley2003,vogeley2004,Karachentsev2004,kreckel_platen2011,kreckel_peebles2011,tikhonov-karachentsev2006,tikhonov2006,karachentsev2006,nevalainen2015}.

The characterisation and study of voids, however, requires surveys of very large volumes measuring large numbers of redshifts (to access large statistics and to resolve in detail properties of the under-dense regions). Forthcoming projects, such as DESI (Dark Energy Spectroscopic Instrument) \citep{desi_pres}, Euclid \citep{euclid_presentaz}, and WFIRST (Wide Field Infrared Survey Telescope) \citep{wfirst_201904} should produce catalogues that are well-suited to this aim, collecting tens of millions of galaxy spectra.  Nearly all-sky photometric surveys like LSST (Large Synoptic Survey Telescope, 2020), will also allow for void measurements using photometric redshifts \citep{pollina2018,calmpit_jain}.

This work is based on the theoretical void size distribution emanating from the Sheth \& van de Weygaert \citep{svdw2004} double barrier excursion set formalism that models the hierarchical evolution of the void population (see also  \citep{vandeWeygaert-bond2008}), more recently extended by Jennings et al. \citep{jennings}. The void size distribution was first measured in \citep{platen2008} in \lcdm simulations using the watershed transform, confirming predictions of \citep{svdw2004} for its shape. Aside from void size function measurements, \citep{platen2008} determined the void shape distribution, and studied alignments and correlations of voids.

The Fisher-matrix approach shows that the number of voids as a function of their size (\ie the {\sl void size distribution function}, also known as {\sl void abundance}) is a promising tool to understand dark energy \cite{pisani2015}. Voids are, thus, complementary to other classical probes of the equation of state (EoS) of dark energy. 
In this paper we extend this investigation further, measuring the void size function from dark matter halo catalogues obtained from the ``Dark Energy and Massive Neutrino Universe'' (DEMNUni) simulation \citep{DEMNUni_paper}, which accounts for different EoS of dynamical dark energy. In particular, we use four models characterised by different combinations of the parameters $w_0$ and $w_a$, introduced in the popular Chevallier-Polarski-Linder (CPL) parametrisation of dynamical dark energy \cite{chevallier,linder}. Direct comparison of these measurements tests if the void abundance is sensitive enough to distinguish different EoS, for an hypothetical survey with the same volume, density, and resolution as our simulations. Comparison to the theoretical description of the void size function \citep{svdw2004,jennings} tests the ability of theoretical models to reproduce the ``observed" measurements and capture the details of the cosmological dependence. We focus our analysis on the simulation boxes centred at $z \sim 1$, corresponding to the typical redshift that next generation surveys will sample.

The paper is organised as follows: in Section \ref{simu_vide} we introduce the simulations; in Section \ref{void_finder} we describe the void finder used to build the void catalogues. Section \ref{theo_sec} presents the theoretical model of the void size function and shows how to self-consistently match it to the simulated void catalogues.  Section \ref{analysis} discusses whether the degeneracies between different dark energy EoS can be lifted using void abundance (Sec. \ref{break_deg}), how variations in the dark energy EoS impact void statistics (Sec. \ref{expected_ev_subsec}), and whether the theoretical predictions match the void size function measured in the simulations (Sec. \ref{matching_theo}). We conclude in Section \ref{conclu}.

\section{Simulations}\label{simu_vide}

Our work is based on the ``Dark Energy and Massive Neutrino Universe'' simulations (DEMNUni) \citep{DEMNUni_paper,DEMNUni_castorina}, performed using the TreePM-SPH code {\sl Gadget-3} \citep{gadget-3_2005}. While the simulation suite also implements massive neutrino effects, in this paper we consider the massless neutrino cases, focusing on variants of the dark energy scenario. Recent companion papers study voids in the massive-neutrino realisations of DEMNUni \citep{kreisch_neutrini2018,schuster2019_neutrino_voids}.

The simulations have a Planck 2013 \citep{planck2013} baseline \lcdm reference cosmology, with $\Omega_\mathrm{m} = 0.32$ and flat geometry. We analyse four dark energy variants, described by the CPL parametrisation \cite{chevallier,linder}
\begin{equation}
w(a) = w_0 + (1-a)w_a \quad \Rightarrow \quad w(z) = w_0 + \frac{z}{1+z} w_a\, ,
\end{equation}
with parameter values chosen within the 2015 Planck boundaries \citep{planck2015_params}: $w_0=[-0.9, \: -1.1]$ and $w_a = [-0.3, \: 0.3]$. In the massless neutrino case considered here, the simulations are characterised by a softening length $\epsilon = 20 \: h^{-1}\mathrm{Kpc}$, and a comoving volume of $(2 \: h^{-1}\mathrm{Gpc})^3$ filled with $2048^3$ dark matter particles with mass $M=8 \times 10^{10} \: h^{-1}M_\odot$. All five numerical simulations (\lcdm plus the four EoS variants) were started at redshift $z=99$, producing 63 different time outputs, logarithmically spaced in the scale factor $a = 1/(1 + z)$, down to $z=0$.  

Voids are identified and selected within dark-matter halo catalogues built using a friends-of-friends (FoF) algorithm \citep{gadget2009,gadget2001}, with minimum halo mass fixed to $M_\mathrm{FoF} \simeq 2.6 \times 10^{12} \: h^{-1}M_{\odot}$ \citep{DEMNUni_paper}. We note that the DEMNUni simulations have a volume and resolution mimicking the data expected from large surveys such as the spectroscopic Euclid wide survey, hence allowing to explore the population of relatively large voids (which are not captured by smaller simulations), and assess their constraining power for cosmology. For future work it would be important to explore the constraining power of the low volume/mass tail of the void size function by analysing smaller simulations reaching a lower minimum halo mass.
In selecting voids from the distribution of halos in different cosmologies, two possible approaches can be followed: (a) selecting all halos down to a fixed mass threshold, or (b) selecting all halos down to a mass when a fixed mean number density threshold is reached. Both methodologies have advantages and disadvantages. For our purpose, we adopt the former approach, \ie select down to a fixed halo mass threshold. This, consequently, results in a different number density of halos when varying the cosmology.   The advantage of calibrating on a halo mass cut (which can be related to a luminosity or stellar mass cut) is that of reproducing what is usually done on observational data, avoiding further sub-sampling to achieve identical number densities, as required by method (b). Clearly, in this case part of the constraining power might be degenerate with the halo mass function (\ie the difference in the void size function could be due to a difference in the halo mass function). We do not investigate here how independent these two effects are in this respect; since systematic errors will be plausibly different for the void mass function, we consider this approach to be more ``agnostic".

\section{Void finder}\label{void_finder}

To identify voids and build the  void catalogues in each simulation we used the ``Void IDentification and Examination'' (VIDE)\footnote{\url{http:www.cosmicvoids.net}} public toolkit \citep{VIDE}.
VIDE is based on the tessellation plus watershed void finding formalism--introduced by Platen et al. (2007) with the Delaunay tessellation/DTFE \citep{schaap2000dtfe} and subsequently used by ZOBOV \citep{neyrinck} with the Voronoi tessellation of the tracer particle population (in our case, dark-matter halos), to estimate the density field based on the underlying particle positions. The algorithm first groups nearby Voronoi cells into zones, corresponding to local catchment ``basins". 
VIDE voids are obtained by merging basins with the watershed transform if the ridge-line separating two basins has a density lower than 20\% of the mean density (see \citep{VIDE}). The value of 20\% is a reference to the isolated spherical void model \citep{platen2007,svdw2004,blumenthal1992} and corresponds to the hypothetical shell-crossing transition of isolated, perfectly spherical voids---that hardly correspond to real voids. The cleaning routine implemented by \citep{ronconi,contarini2019}, and used in our work, is an attempt to reduce the difference between the theoretical isolated, perfectly spherical voids and observed voids (see Sec. \ref{theo_sec}). Future work relying on simulations or survey mocks should thoroughly assess the impact of this threshold choice, as well as the dependence on the shape of voids, the impact of surveys masks, etc. 
Topologically-identified watershed basins and ridge lines are used to construct a nested hierarchy of voids. The algorithm begins identifying the initial zones as the deepest voids, and as it progressively merges voids across ridge-lines, it establishes parent voids and children sub-voids.

\begin{figure}[t]
\centering
\includegraphics[width=\textwidth]{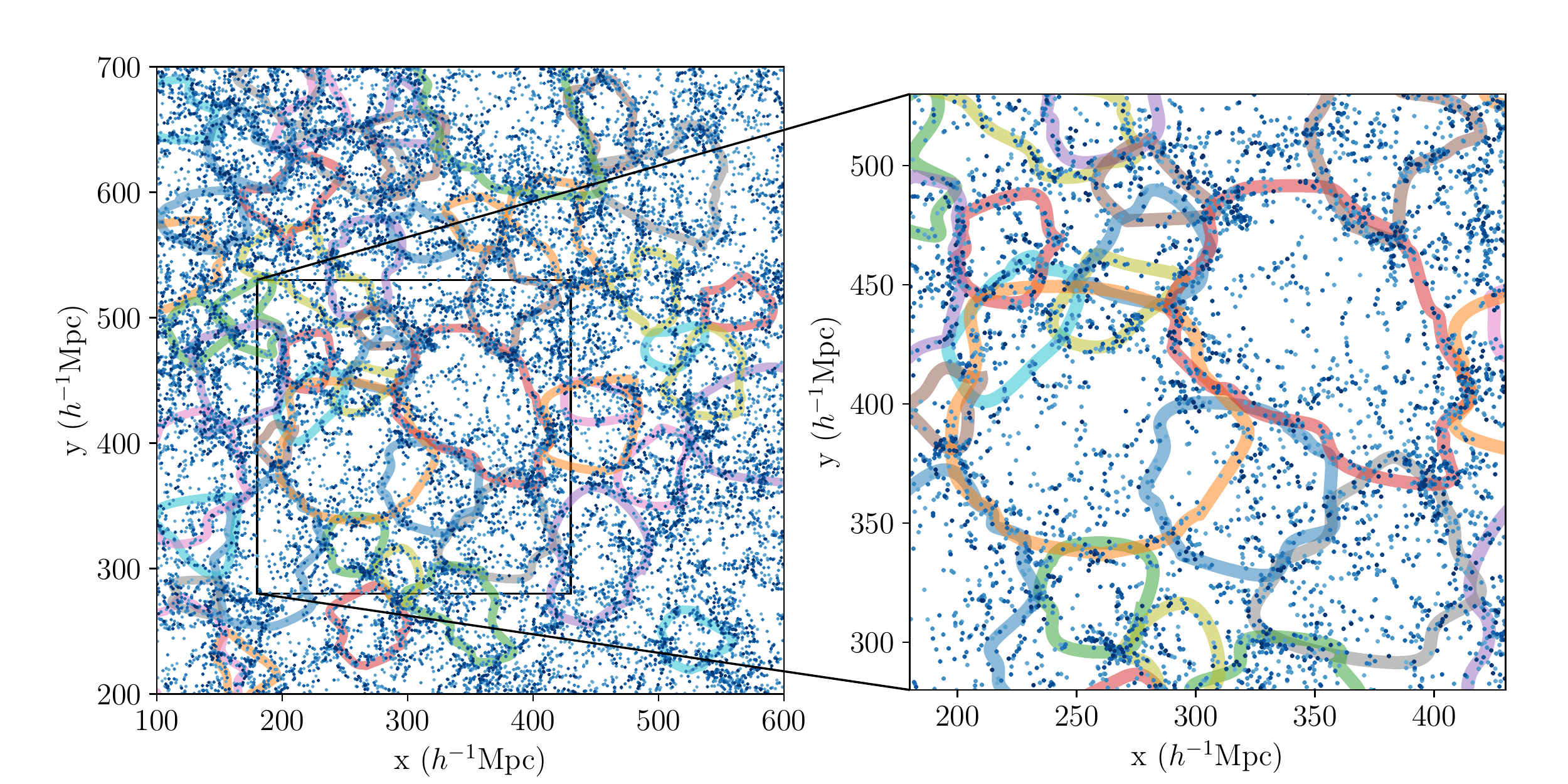}
\caption{Projected void boundaries (colored lines) and halo field (blue dots represent halos, darkness is a function of halo mass) in a slice of the \lcdm simulation at z=1.}
\label{HaloVoidMap}
\end{figure}

VIDE can be launched on any catalogue of tracers, such as dark matter particles and halos (or galaxies) in simulations, as well as real galaxies in surveys; it is also capable to handle a survey selection function and a mask. These features make VIDE a very flexible tool to compare voids in data and simulations, as witnessed by its extensive cosmological applications over the past few years \citep{VIDE,pisani2015,hamaus_constraints_voids,hamaus_rsd_2015,sutter2013aVoid_galaxy,pollina2017,pollina2018}. VIDE provides basic void information, such as their volume $V_{\mathrm{void}}$ (calculated as the sum of the volumes of all contributing Voronoi cells), effective radius, central density, shape via the inertia tensor, eccentricity, etc.; it also yields information about the void hierarchy. One important quantity is the void barycentre, weighted by the volumes of the contributing Voronoi cells: 
\begin{equation}\label{vide_baricentre}
\Vec{X}_v = \frac{1}{\sum_i V_i} \sum_i \vec{x}_i V_i\, ,
\end{equation}
where  $x_i$ and $V_i$ are the positions and Voronoi volumes of each tracer particle $i$ within the void. Such definition implemented in VIDE uses information from the whole object, allowing for a more robust identification of the void centre---where inevitably very few tracers are present, by definition.  A robust centre definition is of the utmost importance: it allows a robust study of the density profile and statistics of  voids, improving the connection between measured and theoretical void abundances (see Section \ref{cleaning_theory_subsec}) \cite{hamaus_density}.

In our analysis we mainly exploit two void features measured by VIDE. The first is the void size, measured as an effective radius: $R_\mathrm{eff} = \left(3V_{\mathrm{void}}/(4 \pi) \right)^{1/3}$, the radius corresponding to a sphere with volume equal to the volume of the void.
The second quantity that is relevant for our analysis is the central density $\rho_\mathrm{central}$, defined as the mean tracer density within a sphere of radius $R_\mathrm{eff}/4$ around the volume-weighted barycenter.

\section{Theory and setup}\label{theo_sec}

In this Section we introduce the theoretical framework to model cosmic void abundances (Sec. \ref{Vdn_theo_subsec}), and discuss how to trim void catalogues and match theory with measurements (Sec. \ref{methodol_subsec}).

\subsection{Theoretical model}\label{Vdn_theo_subsec}

In analogy to the analysis involving collapsed objects, it is possible to predict theoretically the number of voids per unit volume, as a function of cosmological parameters. A method widely used in the literature is an extension to voids of the excursion set model, developed within the framework of the halo mass function \citep{press_schecther,sheth_tormen1999,sheth_tormen2001,sheth_tormen2002}, to obtain the distribution function of void sizes. First proposed by Sheth \& van de Weygaert \citep{svdw2004} (SvdW) (see also \citep{paranjape2012}), this approach has been more recently extended by Jennings et al. \citep{jennings}.
Its physical rationale is based on the realisation that voids evolve hierarchically. The model entails an idealised mathematical description of this process (see \citep{sahni1994,aragonCalvo2010,aragonCalvo2013}). The idea of using the excursion set model is based on the linear statistical properties of the matter-field perturbations and on the spherical nonlinear collapse model. Contrary to the halo abundance, which relies on halo masses, void abundance is described in terms of void radii. This is a proxy to the void volume, which is a more directly observable quantity for voids, \eg compared to the halo mass.
To study the abundance of cosmic voids we make two strong approximations: first, we assume a perfectly spherical nature for voids, second we consider that voids are isolated objects (see \eg \citep{vandeWeygaert2016} for a discussion). While an isolated void may be argued to tend towards sphericity in its evolution \citep{icke1984}, Platen et al. \citep{platen2008} showed that real voids behave differently, in particular the small ones ($R_\mathrm{eff} < 10 h^{-1} \mathrm{Mpc}$). Even though for collapsing objects an elliptical model is required \citep{sheth_tormen1999,sheth_tormen2001,sheth_tormen2002}, and despite the fact that for small voids the effects of the environment need to be better understood \citep{platen2008,vandeWeygaert2016}, the spherical approximation is accurate enough to obtain a reliable void size function for relatively large voids (see also the discussion in Sec. \ref{cleaning_theory_subsec}). In particular, this approximation appears to be even more accurate in the inner regions of voids---on which our work is focused: in the central parts of voids the effects of the environment are somewhat suppressed, allowing for an increased evolution towards sphericity \citep{vandeWeygaert1993}.

The void size function depends on the top-hat model, according to which a void is a spherical object in the matter distribution, with mean density contrast equal to a fixed negative value. The mean density contrast within the radius $r$ of a shell is:
\begin{equation}\label{dens_contr_eq}
\Delta (r) = \frac{3}{r^3} \int_0^r \left[ \frac{\rho(r')}{\langle \rho \rangle} - 1 \right] r'^2 \text{d}r' \, ,
\end{equation}
where $\langle \rho \rangle$ is the mean density of the Universe.
According to the top-hat model, overdensities experience two main events during their evolution, \textit{turn-around} and \textit{virialization}. On the contrary, the top-hat evolution of underdensities is not characterised by any specific event, they continue their outward-directed expansion forever. Therefore it is common to consider the \textit{shell-crossing} condition as the event that characterises the void formation \citep{bertschinger1985,blumenthal1992,svdw2004,jennings,platen2007,dubinski1993}. This condition strictly depends on the initial density profile of the underdensity.
For a top-hat initial density profile, it happens at a defined value of the density contrast: $\Delta_\mathrm{SC} \simeq -0.7953 $ in an Einstein-de Sitter universe. Importantly, the properties of linear cosmological fluctuations, on which we rely to build the theoretical void size functions, do not depend on any specific underdensity threshold nor on the shell-crossing condition. In principle, we can calculate void size functions for any negative threshold value, not only for the one corresponding to shell-crossing.  This feature makes the void size function particularly versatile.
We can express the abundances of voids, or more generally of fluctuations in the matter field, as a function of their mass as \citep{svdw2004,jennings}:
\begin{equation}\label{massVoid_func}
\frac{\text{d} n_\mathrm{L}}{\text{d\,ln } M} = \frac{\langle \rho \rangle}{M} f_{\text{ln } \sigma}(\sigma) \frac{\text{d\,ln } \sigma^{-1}}{\text{d\,ln } M}\, ,
\end{equation}
where $f_{\text{ln } \sigma}(\sigma)$ is the fraction of fluctuations that become voids. To calculate this quantity we use the excursion set formalism with two density thresholds: one for void formation $\delta_\mathrm{v}$, and one for collapse $\delta_c \simeq 1.686$. The threshold for collapse is necessary to take into account the fraction of underdensities that lie within larger overdensities, \ie the void-in-cloud fraction. This class of underdensities will be squeezed out of existence by the larger collapsing overdensity that surrounds them \citep{svdw2004}. Hence, in order to evaluate the fraction of fluctuations that will evolve in voids, we calculate the probability
that a fluctuation of scale $R$ (\ie the fraction of random walks that) crosses $\delta_\mathrm{v}$, the void formation threshold,
and have never crossed $\delta_\mathrm{c}$, the threshold for collapse, at any scale larger than $R$:
\begin{equation}
f_{\text{ln } \sigma}(\sigma) = 2 \sum_{j=1}^{\infty} e^{-\frac{(j \pi x)^2}{2}} j \pi x^2 \sin{\left( j \pi \mathcal{D} \right)}\, ,
\end{equation}
with
\begin{equation}
\mathcal{D} = \frac{|\delta_\mathrm{v}|}{\delta_c + |\delta_\mathrm{v}|} \qquad x = \frac{\mathcal{D}}{|\delta_\mathrm{v}|} \sigma \, ,
\end{equation}
where $\sigma$ is the variance of linear matter perturbations, and all the quantities are computed in the linear regime, on which the excursion set formalism relies \citep{jennings}.

In order to connect the theory with observable voids we have to convert linear quantities into the corresponding nonlinear ones. To obtain the number density of voids as a function of their size, a more relevant quantity for voids, we convert, in the linear regime, the mass $M$ in Eq. (\ref{massVoid_func}) into the volume: $M \rightarrow V(r_L)$. As in Jennings et al. \citep{jennings} we impose ``volume conservation'', \ie we fix the void volume fraction of the Universe in linear regime to be equal to the one in the nonlinear regime:
\begin{equation}\label{volume_cons}
V(r) \text{d} n = V(r_L) \text{d} n_L\rvert_{r_L = r_L(r)}\, ,
\end{equation}
where $r_L$ indicates the shell radius in the limit of linear perturbation theory, and $r$ the nonlinear shell radius. The conversion for the linear to the nonlinear shell radius is given by the evolution of perturbations in nonlinear regime
\begin{equation}\label{r_r_L}
\frac{r}{r_L} = \left( \frac{\langle \rho \rangle}{\rho_v} \right)^{1/3}\, .
\end{equation}
Eq.(\ref{volume_cons}) accounts for the fact that void merging preserves the volume fraction of voids in the Universe, hence it accounts for void evolution considering that voids are not isolated objects. Substituting in Eq. (\ref{massVoid_func}) the condition given by Eq. (\ref{volume_cons}), together with the conversion $M \rightarrow V(r_L)$, we obtain \citep{jennings}:
\begin{equation}
\frac{\text{d} n}{\text{d\,ln } r} = \frac{f_{\text{ln } \sigma}(\sigma)}{V(r)} \frac{\text{d\,ln } \sigma^{-1}}{\text{d\,ln } r_L} \frac{\text{d\,ln } r_L}{\text{d\,ln } r}\biggr\rvert_{r_L = r_L(r)}\, .
\end{equation}
While the fraction $\rho_v/ \langle \rho \rangle$ of Eq. (\ref{r_r_L}) could in principle be written as a function of the void shape, we describe voids as spherical. So this ratio is the same for each void, \ie $\text{d\,ln } r_L / \text{d\,ln } r = 1$, and we can rewrite the void size function as
\begin{equation}
\frac{\text{d} n}{\text{d\,ln } r} = \frac{f_{\text{ln } \sigma}(\sigma)}{V(r)} \frac{\text{d\,ln } \sigma^{-1}}{\text{d\,ln } r_L} \biggr\rvert_{r_L = r_L(r)}\, .
\end{equation}
Due to the condition of Eq. (\ref{volume_cons}), this void size function is known as the ``volume conserving model'', or Vdn. It corresponds to the SvdW model with the additional condition of conserving volume, which basically shifts the void size function towards lower values.

\subsection{Analysis setup}\label{methodol_subsec}

In this Section we first explain how to select voids that are reliable and provide a high signal-to-noise ratio (Sec. \ref{selectig_voids_subsec}). In (Sec. \ref{cleaning_theory_subsec}) we then discuss how to prepare void catalogues from simulations to allow a comparison with the theoretical model. These steps are necessary to match the theoretical forecast of void abundance with simulations or data.

\subsubsection{Criteria to select voids}\label{selectig_voids_subsec}

To test the sensitivity of the void size distribution to the dark energy EoS, we first compare empirically the abundances measured in the simulations. The pipeline/toolkit we used is VIDE, which is built around the parameter free watershed transform void finding formalism (see \citep{platen2007,neyrinck,roerdink2000} for the watershed transform), hence including all the underdensities found in a given dataset. We thus explore whether filtering the void catalogues according to suitable void physical features could enhance the sensitivity to the EoS. We follow two criteria. 

The first criterion aims at minimising systematic effects, such as noise and the fraction of spurious voids (\eg Poisson voids \citep{neyrinck}). The mean halo density in the simulations fixes the spatial resolution of the halo catalogues; thus, we discard all voids that have an effective radius smaller than the mean halo separation, defined as $R_\mathrm{mhs}=\left[ N_\mathrm{halos}/V_\mathrm{sim} \right]^{-1/3}$. In each DEMNUni box at $z=1$ used in this work we have $\sim 11 \times 10^6$ halos and $R_\mathrm{mhs}$ is approximately 9 \hmpc. 

The second criterion aims at maximising the signal-to-noise ratio, using the void central density. With the goal of improving the match between observed and modelled voids, Jennings et al. \citep{jennings} suggested to filter the void catalogue according to their core density, \ie the density of the central Voronoi cell used in the void definition. Here we apply a similar technique, but using the void central density and then check the consistency of the results both with and without filtering. In this way, we select voids that share similar features, \eg they have a depth below a fixed value. 

Therefore we select voids with $\rho_\mathrm{central}=0$, \ie voids with no halo within $R_\mathrm{eff}/4$. This may seem rather artificial, since we know that void galaxies can be found in the central regions of voids (\eg in the SDSS galaxy survey; see \citep{kreckel_peebles2011,kreckel_platen2011}), in fact it simply implies that we exclude voids with no galaxies below a certain mass in their central region. The strength of this filter depends on the tracer we use, that is, in our case, on the minimum halo mass of the simulations ($\sim 2.6 \times 10^{12} h^{-1}M_\odot$). We checked that choosing a  milder filtering does not change the overall results. Hereafter we will refer to the catalogues selected with the $\rho_\mathrm{central}=0$ threshold as the ``filtered catalogues".

In Fig. \ref{VIDE_all} we present the first results of our analysis (see also Sec. \ref{break_deg} for a discussion), \ie the void abundances as measured from the filtered (right panel) and unfiltered (left panel) void catalogues extracted from the DEMNUni simulation. We note that large voids are more sensitive to the central density threshold cut; when the density threshold is lowered they are the first to decrease in number. We discuss the mechanism producing this behaviour. If we consider the life of a single void, during its evolution it increases in size and becomes deeper in the inner regions, due to matter evacuation (see \eg Fig. 3 of \cite{svdw2004}). This dynamics suggests that large voids are also more evolved. On the other hand, we expect also that large voids begin their evolution as large shallow underdensities, since, given the shape of the matter power spectrum, fluctuations on larger scales have lower variance. As such, it is statistically rare to find large voids that are also deep. It follows that, considering families of voids with the same age but different sizes, on average large voids are shallower than small ones. Fig. 1 in Hamaus et al. \citep{hamaus_density} confirms this scenario.

\subsubsection{Void catalogue preparation}\label{cleaning_theory_subsec}

The theoretical definition of a void as a dark matter underdensity hardly matches the object identified in observations, \eg the watershed void found in the distribution of galaxy tracers \citep{furlanetto_piran}. The theoretical void size function relies, in fact, on the top-hat model, according to which a void is a spherical object defined by a mean density contrast equal to a given (negative) value Eq. (\ref{dens_contr_eq}). To be able to compare observational data with theory, therefore, we have to be able to reconcile these different void descriptions.

A first method to connect theory and observations is to leave $\delta_\mathrm{v}$ as a free parameter and calibrate it in order to make the theoretical void size functions to fit the measured void abundance in the considered data set. This method allows to overcome the problem by parametrising our ignorance \citep{furlanetto_piran, pisani2015,chan_voidClustering_2014}. On the other hand, as suggested by Jennings et al. \citep{jennings}, it would be interesting to directly connect the top-hat definition to observed voids, matching the theoretical void size function with the measured abundances in the data set \textit{without any free parameter}.
This second option is more attractive, we thus implement it for the given DEMNUni void catalogues.

The most direct way to connect the top-hat void definition with voids found by VIDE is to reconstruct their density profile using the spherical shape approximation. This is possible for three main reasons. First, void ellipticity is low: for example, in the \lcdm case we measure a mean ellipticity $e=0.13^{+0.06}_{-0.05}$. We note that the low ellipticity is verified only for relatively large voids, such as the ones we consider in this work. Small voids are considerably more aspherical as they are more strongly influenced by external environmental effects such as large scale tidal forces, are embedded in collapsing overdensities, and are squeezed against neighbouring voids \citep{platen2008}---they would need a more careful consideration, that we leave for future work. Second, with our large volumes and therefore large statistics, voids can be safely considered as spherical on average \cite{lavaux2012_APtest}. Finally, the abundance is a statistical measure for which in any case individual void features are averaged out. Summarising, to first approximation, in deriving the void abundance we can consider voids as spherical objects.  This, together with the fact that void profiles are universal \citep{hamaus_wandelt2014}, gives us a way to connect our VIDE voids with the top-hat model definition. 

We thus find the radius $r$ inside which the mean density contrast equals the chosen threshold. This is the crucial point in the procedure, as it re-normalises the void radii to the same density contrast. We note that to reconstruct the spherically averaged void density profile a robust void centre definition is crucial. Therefore relying on a void centre found from a non-spherical void-finding algorithm enhances the reliability of the methodology. This is the case for the volume-weighted barycentre of VIDE defined in Eq. (\ref{vide_baricentre}). In contrast, void centre definitions based on sparse particle counts can be more affected by sampling artefacts and less optimal for this application. 

We then check whether void abundances obtained with resized voids match the theoretical void size functions. Following \cite{jennings}, we adopt the following strategy:

\begin{itemize}
\item We resize all VIDE voids to the fixed threshold, which we call $\delta_\mathrm{v,NL}^\mathrm{H}$, where $H$ stands for halos, the tracer using for the void finding procedure. We apply the resizing using the code by Ronconi \& Marulli \citep{ronconi}, available in the CosmoBolognaLib C++/Python library\footnote{\url{https://github.com/federicomarulli/CosmoBolognaLib}} \citep{cosmobolognalib}.  The subscript NL (Non Linear), here and in the following, indicates when we are considering fully nonlinear quantities. 

\item The theoretical void size function is referred to the statistical properties of matter, not of the tracers. We follow Pollina et al. \citep{pollina2017,pollina2018} to compute the threshold in the matter $\delta_\mathrm{v,NL}^\mathrm{m}$, which corresponds to the one in the halo distribution and then invert the linear bias relation $\delta_\mathrm{v,NL}^H=b_\mathrm{eff} \times \delta_\mathrm{v,NL}^\mathrm{m}$ (with $b_\mathrm{eff}$ being the tracer bias). 

\item In order to calculate the theoretical void size function, we need to express the matter underdensity threshold in linear theory. This is obtained in the spherical collapse model by converting the nonlinear density contrast, $\delta_\mathrm{v,NL}^\mathrm{m}$, into the corresponding linear one, $\delta_\mathrm{v}$.\footnote{In EdS the inversion can be done analytically, being 
$\delta_\mathrm{v,NL}^\mathrm{m}(\eta)=\frac{9}{2} \frac{(\sinh{\eta} - \eta)^2}{(\cosh{\eta}-1)^3}$, while the corresponding linear contrast is $\delta_\mathrm{v}(\eta) = \frac{3}{20}[6(\sinh{\eta}-\eta)]^{2/3}$. In the spherical evolution model, the value of $\eta$ corresponding to the full nonlinear quantity $\delta_\mathrm{v,NL}^\mathrm{m}$ allows us to recover the corresponding linear theory underdensity $\delta_\mathrm{v}$.} 
We verified that this conversion is weakly dependent on the redshift and variations of the dark energy EoS. This would allow marginalisation 
over $\delta_\mathrm{v}$, when considering the different EoS implemented in the DEMNUni simulations (as done in \cite{pisani2015}).

\item To account for the redshift dependence of the  void size function, we extrapolated all quantities using linear theory to the present time, $z=0$, \citep{sheth_tormen2001}, as it is done in the case of the halo mass function \citep{svdw2004,sheth_tormen2002}:
\begin{equation}
\begin{split}
\delta_\mathrm{v} &\rightarrow \delta_\mathrm{v}/D(z) \\
\delta_c &\rightarrow \delta_c/D(z) \\
\sigma &\rightarrow \sigma(z=0)\, ,
\end{split}
\end{equation}
where $D(z)$ is the growth factor in linear theory \citep{percival2005} normalised to unity at z=0. 
\end{itemize}
We now have all the tools to calculate the theoretical void size function and compare it with the void abundances of the resized void catalogues from the simulations.

\section{Analysis}\label{analysis}

In this section we first compare the measured size functions from the different simulations, discussing how sensitive they are to variations of  the dark energy EoS. Secondly, we verify whether the theoretical model is capable to match the corresponding simulation.

\subsection{Sensitivity to dark energy EoS}\label{break_deg}

While we can distinguish different dark energy EoS from \lcdm using traditional observables (such as the halo mass function), void abundances can contribute in breaking the degeneracy between different models as they are subject to different systematics and are potentially more sensitive since they are dark-energy dominated objects. We first consider the total number of voids in each cosmology (a simpler observable), and then analyse in more detail the overall shape of the void size function.

As shown in Table \ref{TotNumCosm}, by simply measuring the total number of simulated voids for each dark energy model, we can break the degeneracy between the \lcdm model and the EoS with $[w_0, w_a]=[-0.9,0.3]$ and $[w_0, w_a]=[-1.1,-0.3]$. This is coherent with \cite{pisani2015}, given the Fisher matrix orientation, and confirms the naive expectation that it is easier to distinguish CPL models where $w(z)$ is farther from \lcdm.

\begin{figure}[t]
\centering
\includegraphics[width=\textwidth]{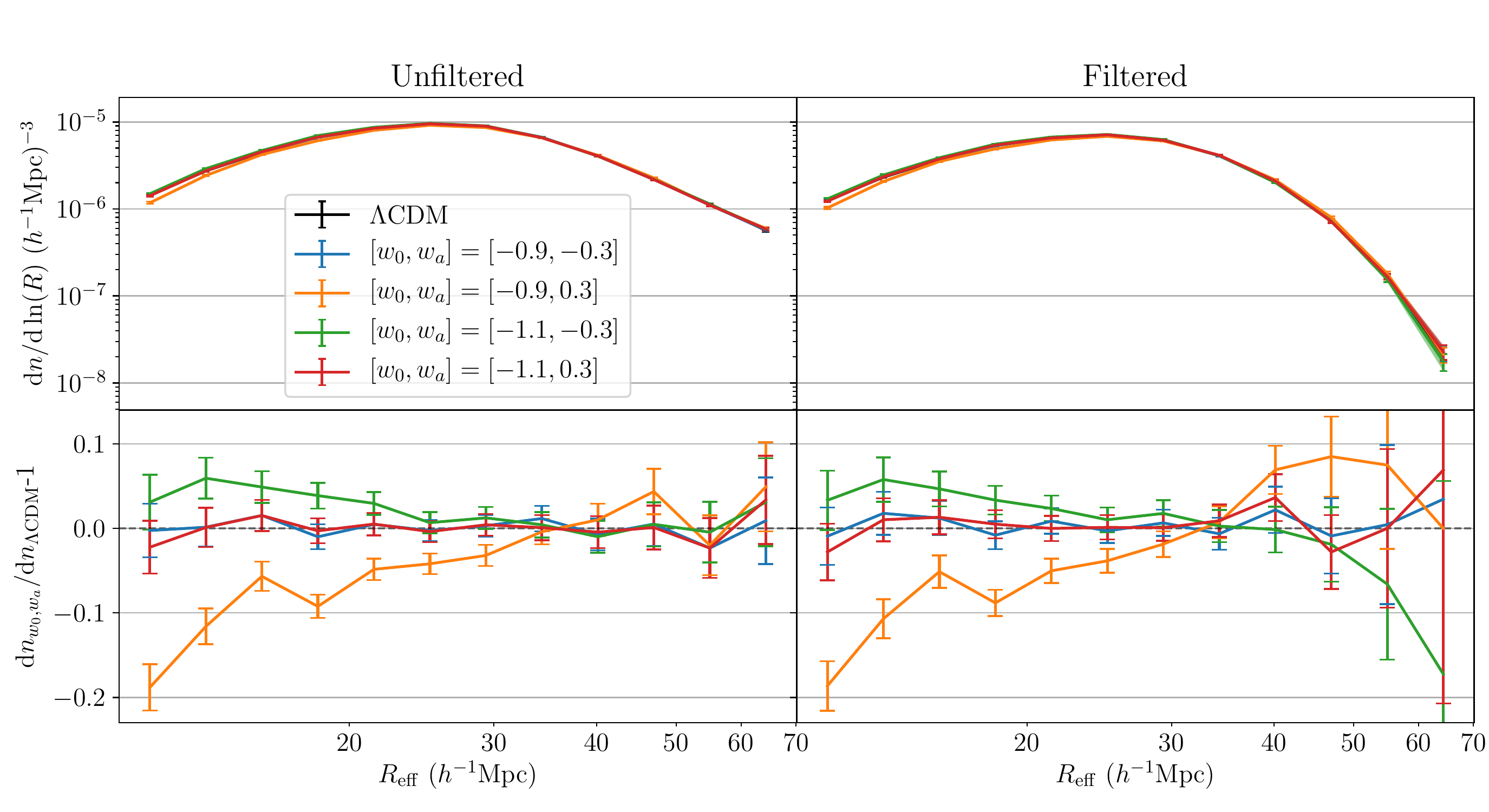}
\caption{Void abundances measured in the DEMNUni simulations without any void selection (left) and filtered (right). The upper charts show abundances for the five different cosmologies. The lower charts show the relative abundances against the \lcdm case. The errorbars are the Poissonian uncertainties.}
\label{VIDE_all}
\end{figure}

As mentioned above, in Fig. \ref{VIDE_all} we compare the full void size function for the four evolving dark energy EoS and \lcdm. For small radius values, the void size functions with $[w_0, w_a]=[-0.9,0.3]$ and $[-1.1,-0.3]$ are clearly distinguishable from \lcdm. Conversely, at large radii Poisson noise dominates. The filter mildly increases the separation from \lcdm at larger radii: while its effect is stronger for the $[w_0, w_a]=[-0.9,0.3]$ case (orange curve), for the other models uncertainties remain high and filtering only provides a mild enhancement to break degeneracy. A larger volume will increase the statistic of large voids, promising to make the filter more effective.

This analysis shows that, for a survey with volume of 8 $(h^{-1} \mathrm{Gpc})^3$ at $\langle z \rangle = 1$, measuring the void size function allows to disentangle \lcdm from a $[w_0, w_a]=[-0.9,0.3]$ dynamical dark energy model at the $8.9\sigma$ confidence level, and the $[-1.1,-0.3]$ model at a $4\sigma$ confidence level.  Milder deviations (such as models with $[w_0, w_a]=[-0.9,-0.3]$ or $[-1.1,0.3]$), are instead marginally distinguishable within our simulated volumes using this statistics. Considering a bin size of $\delta z = \pm 0.1$ centred at $z = 1$, a survey such as Euclid will be characterised by a volume close to DEMNUni. Of course considering the whole light-cone would provide a much larger volume, implying a stronger constraining power (in addition to exploiting redshift dependence).

We note that the use of a $0^\mathrm{th}$ order Voronoi tessellation based density field does not allow to robustly identify significant small voids probed by only a few halos, because the flat density profile in a Voronoi cell involves a considerably stronger noisy density field (with accompanying deviations) than, for example, the use of the first-order Delaunay tessellation density field \citep{vandeWeygaert2007}. It is also to reduce the impact of this issue that we chose to exclude small voids from our analysis---which bears additional advantages since environmental effects are expected to be stronger at those smaller scales. We note that we excluded voids with radii below one time the mean halo separation for the analysis, and the resizing technique additionally removes voids with resized radii below twice the mean particle separation (see Sec. \ref{matching_theo}).

While degeneracy breaking is a first step, building a reliable theoretical prediction of the abundance in other cosmologies is far more powerful to constrain parameters. We first discuss in the next Section an interpretation of measured abundances in simulations, and then focus in Section \ref{matching_theo} on their match with theoretical prescriptions.

\begin{table}[t]
\centering
\begin{tabular}{| c | c | c | c | c |}
\cline{2-5}
\multicolumn{1}{c}{} & \multicolumn{2}{|c|}{\textbf{Unfiltered}} & \multicolumn{2}{c|}{\textbf{Filtered}} \\
\hline
Cosmology & $N_\mathrm{tot}$ & $N_\mathrm{tot}/N_{\mathrm{tot}, \Lambda \mathrm{CDM}} - 1 \: (\%)$ & $N_\mathrm{tot}$ & $N_\mathrm{tot}/N_{\mathrm{tot}, \Lambda \mathrm{CDM}} - 1 \: (\%)$ \\
\hline
$\Lambda \mathrm{CDM}$ & 78589 & --- & 54394 & --- \\
\hline
$[-0.9,-0.3]$ & 78756 & $0.21\pm0.51$ & 54604 & $0.39\pm0.61$ \\
\hline
$[-0.9,0.3]$ & 75169 & $-4.35\pm0.49$ & 42123 & $-4.18\pm0.59$\\
\hline
$[-1.1,-0.3]$ & 80200 & $2.05\pm0.51$ & 55637 & $2.29\pm0.62$ \\
\hline
$[-1.1,0.3]$ & 78658 & $-0.09\pm0.5$ & 54639 & $0.45\pm0.61$ \\
\hline
\end{tabular}
\caption{Total number of voids ($N_\mathrm{tot}$) and relative number of voids with respect to the \lcdm case ($N_\mathrm{tot}/N_{\mathrm{tot}, \Lambda \mathrm{CDM}} - 1$) in the unfiltered and filtered catalogues at $z=1.05$. The uncertainty is Poissonian.}
\label{TotNumCosm}
\end{table}

\subsection{Phenomenology of dark energy effects}\label{expected_ev_subsec}

\begin{figure}[t]
\centering
\includegraphics[width=\textwidth]{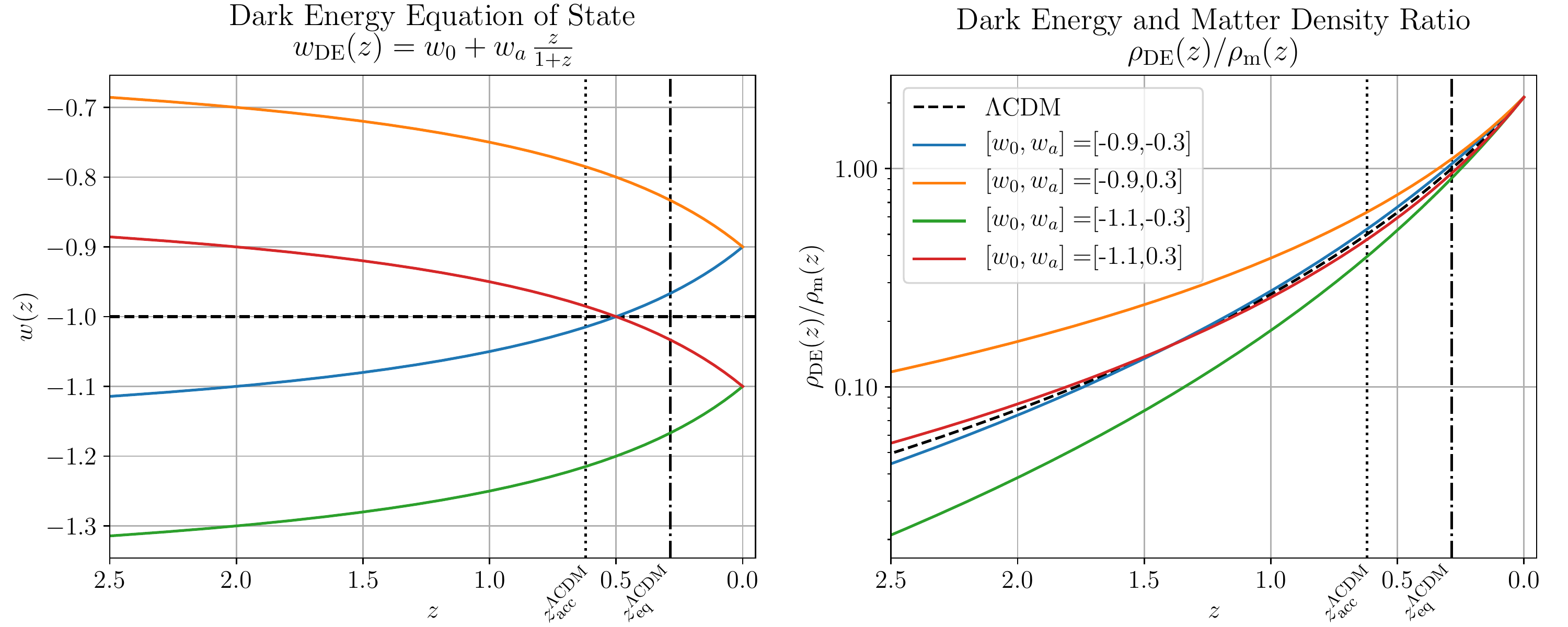}
\caption{Evolution of the dark energy EoS (left panel) and of the dark energy density $\rho_\mathrm{DE}$ (right panel), both as a function of redshift $z$. The parameter $\rho_\mathrm{m}$ is the matter density. In both panels the vertical lines represent respectively the redshift value at which the Universe begins to accelerate its expansion (dotted line, $z_\mathrm{acc}^{\Lambda \mathrm{CDM}}=0.620$), and the redshift of the matter-dark energy density equality (dash dotted line, $z_\mathrm{eq}^{\Lambda \mathrm{CDM}}=0.286$), both for the \lcdm model.}
\label{DE_EOS_evolution_and_values}
\end{figure}

In this section we qualitatively discuss the signature of dark energy with different EoS on void populations, by recognising the physics that comes into play. We aim to show that the ordering of void abundances from the DEMNUni simulations observed in Section \ref{break_deg} matches our physical expectations, focusing on small and large size voids. We recall that the dark energy density value is given by $\Omega_\mathrm{DE} = 0.68$ at $z=0$ for all the simulated cosmologies. Qualitative statements therefore refer exclusively to mild deviations of the dark energy EoS with respect to \lcdm.

The impact of dark energy on voids needs to be analysed considering two aspects: on the one hand, the moment when dark energy becomes the dominant component (hence the fact that its impact becomes relevant earlier or later with respect to \lcdm); on the other hand, the depth of the perturbation, which can be related to its evolution time. To help the discussion, Fig. \ref{DE_EOS_evolution_and_values} shows the evolution of the dark energy EoS and of the dark energy density $\rho_\mathrm{DE}$ as functions of redshift.

As mentioned above, large voids begin their evolution as large and shallow underdensities in the matter field, since the variance of the matter perturbations at large scales is small. It is very unlikely that large voids belong to the void-in-cloud class, hence their evolution is straightforward: they continuously expand. The evolving time in which large voids become deeper voids is roughly comparable with the Hubble time \citep{sutter2014b,wojtak2016}. Therefore the main aspect impacting the evolution of large voids is represented by the time at which the dark energy contribution starts, later or earlier for different EoS  with respect to \lcdm implementation. More precisely, as shown in Fig. \ref{DE_EOS_evolution_and_values}, a dark energy with more negative $w(z)$ dominates later than in \lcdm, and a dark energy with less negative $w(z)$ dominates earlier. Dark energy will have respectively less (or more) time to contribute to the expansion of voids, according to the more (or less) negative value for the EoS. 

For small voids the situation is different: at small scales the variance of the matter field perturbations is large. The underdensities span a large range of depths. In general, according to the top-hat model, the deeper the voids are, the shorter the void evolution time is. The time-scale of the evolution of small voids depends on the particular value of their underdensity in each case, but globally the population of small voids evolves on a time scale smaller than the Hubble time. 
If the dark energy density dominates for a longer time than for \lcdm ($w(z)>-1$), the growth of late time perturbations is slowed down and prevents the formation of new smaller voids. However, large voids have already formed earlier and are now expanding even faster than in \lcdm. Vice versa, if dark energy dominates for a shorter time ($w(z)<-1$), late-time perturbations can grow larger and form more small voids, but the earlier existing large voids expand more slowly than in \lcdm. Referring to our cases: dark energy with $w<-1$ (green curve, left panel of Fig. \ref{DE_EOS_evolution_and_values} at $z=1$) shows more small voids than in \lcdm at the same epoch. Vice versa, at fixed $z$, for $w>-1$ (yellow curve, left panel of Fig. \ref{DE_EOS_evolution_and_values}) there are fewer small voids than in the \lcdm case.

The above reasoning explains qualitatively the behaviour of the void population for $[w_0,w_a]=[-1.1,-0.3]$ and $[w_0,w_a]=[-0.9,0.3]$. 
The other two cosmologies are more entangled with the \lcdm model, as $w_a$  compensates the impact of $w_0$. For these cases the epoch of dark energy domination is slightly affected and so also the abundance of large voids.
While the simplistic discussion above can give an indication of the trend (in particular for small voids) it does not allow us to reasonably predict the behaviour of their abundance. 
Indeed our reasoning focused on the role of dark energy versus interior (under)density in voids, however the dynamics of small voids is largely influenced by their environment \citep{platen2008,vandeWeygaert2016}. Small voids are in fact mostly located near the boundaries of large voids, near the surrounding filaments and walls, and are heavily squeezed by the combined effect of the more prominent expansion of their large neighbour(s) and the gravitational influence of the high-density filaments and walls. While the smallest voids (for which these effects are expected to be stronger) are already excluded from our analysis, for voids of smaller size (among the ones we consider) such simplistic reasoning might fail. Hence, simulation modelling is the best tool to assess the consequences of small voids' behaviour for precision cosmology applications. Future work relying on higher resolution simulations allowing to reach smaller halo mass will be better placed to thoroughly assess the environmental impact.

Table \ref{DE_allvoids} schematically sums up the discussion to explain abundance behaviour for large and small voids for the different EoS compared to the \lcdm case. 

\begin{table}[h]
\centering
\begin{tabular}{| c | c | c | c | c |c|}
\hline
$w_0$ & $w_a$ & DE domination period & Large voids& Global EoS value ($z=1$) & Small voids \\
\hline
\fcolorbox{olive}{white}{-1.1}& \fcolorbox{olive}{white}{-0.3} & shorter than \lcdm&  $\searrow$ & more negative than \lcdm  & $\nearrow$\\
\hline
\fcolorbox{red}{white}{-1.1}& \fcolorbox{red}{white}{~0.3} & shorter than \lcdm & $\sim$ compensated& less negative than \lcdm  & $\sim$ compensated \\
\hline
\fcolorbox{blue}{white}{-0.9}& \fcolorbox{blue}{white}{-0.3}  & larger than \lcdm & $\sim$ compensated& more negative than \lcdm  & $\sim$ compensated \\
\hline
\fcolorbox{orange}{white}{-0.9}& \fcolorbox{orange}{white}{~0.3} & larger than \lcdm & $\nearrow$& less negative than \lcdm  & $\searrow$\\
\hline 
\end{tabular}
\caption{Qualitative description of the impact of the considered dark energy EoS on the population of voids with respect to the \lcdm case ($[w_0,w_a]=[-1,0]$). The arrows indicate if the population is increased ($\nearrow$) or reduced ($\searrow$) with respect to the \lcdm case, and colours correspond to the plot.}
\label{DE_allvoids}
\end{table}

\subsection{Theory description}\label{matching_theo}

In this Section we proceed to match the theoretical predictions for void abundances with measurements from the simulations, without resorting to the addition of any free parameter.

\subsubsection{Void abundance match: \lcdm}

\begin{figure}[t]
\centering
\includegraphics[width=\textwidth]{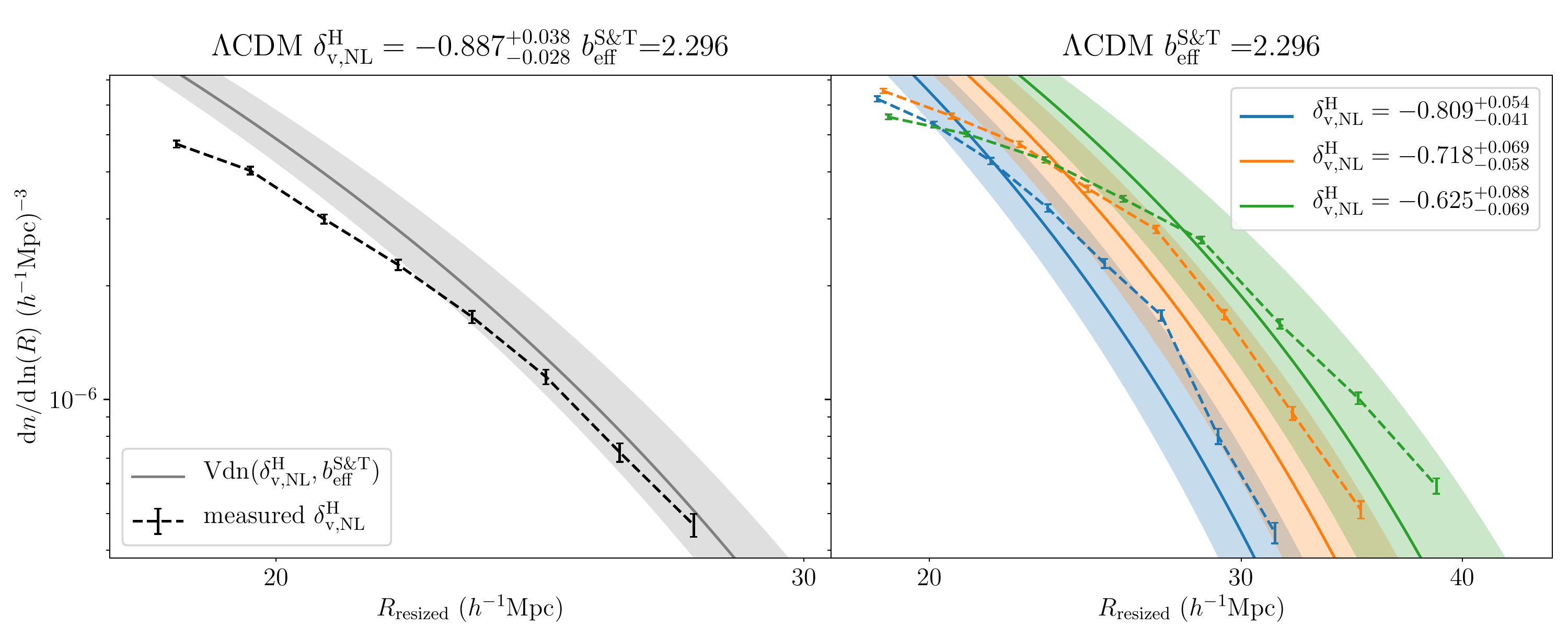}
\caption{Match of the measured abundances of voids (after resizing procedure) with theoretical predictions. The left panel shows the measurement for $\delta_\mathrm{v,NL}^\mathrm{H}=-0.887$ (black curve), and the corresponding theoretical void size function obtained using $b_\mathrm{eff}$ as described in the text (grey curve). The right panel shows the measured void abundances for various thresholds $\delta_\mathrm{v,NL}^\mathrm{H}$ (dashed curves), and the corresponding theoretical void size functions (solid curves). The current resized voids correspond to observable voids in the range $R_\mathrm{eff} = [35-70] \: h^{-1}\mathrm{Mpc}$. Shaded areas give the uncertainty in the resizing procedure (see text).}
\label{Vdn_lcdm}
\end{figure}

We follow the strategy described in Sec. \ref{cleaning_theory_subsec}: we obtain the linear threshold $\delta_\mathrm{v}$ corresponding to the nonlinear threshold in the matter distribution $\delta_\mathrm{v,NL}^\mathrm{m}$; obtained by inverting the linear bias relation:  $\delta_\mathrm{v,NL}^H = b_\mathrm{eff} \times \delta_\mathrm{v,NL}^\mathrm{m}$, and considering the conversion between linear and nonlinear.  The effective bias $b_\mathrm{eff}$ of the FoF halos is computed using the Sheth \& Tormen \citep{sheth_tormen1999} model.
We study the corresponding void size function for the \lcdm case for a fixed $\delta_\mathrm{v}$ value and show our results in Fig. \ref{Vdn_lcdm}\footnote{We note that two recent papers \cite{ronconi2019,contarini2019} tested the same theoretical model on \lcdm halos from simulations, albeit with smaller volume. Aside from looking at different EoS for dark energy in our paper, we confirm the validity of Vdn model in the \lcdm case with DEMNUni simulations, much larger in volume and with about the same mass resolution.}. In the left panel, the black dashed curve represents the measured void abundance of the resized void catalogue for \lcdm, while the grey curve represents the theoretical void size function for one fixed $\delta_\mathrm{v}$ value in the halo distribution with $b_\mathrm{eff}$ the effective halo bias. We note that the radius range of this analysis is different from the one shown in Sec. \ref{break_deg}, where the effective void radius was defined via the watershed approach. Here the radius $R_\mathrm{resized}$ refers to spherically resized voids, and therefore is smaller than the effective radius of watershed voids. The resized void range corresponds to observable voids in the range $R_\mathrm{eff} = [35-70] \: h^{-1}\mathrm{Mpc}$.

The uncertainty in the theoretical void size functions is due to the resizing procedure and is estimated in the following way. We fix the threshold value $\delta_\mathrm{v,NL}^\mathrm{H}$ we wish to consider and we re-scale each void up to the corresponding radius using the algorithm in \cite{ronconi}. We then calculate independently the actual density contrast within the re-scaled radius for each void, we use the peak of the distribution of the measured density contrast as the effective threshold value, and quantify the uncertainty in the resizing procedure via the half-width-half-maximum of the measured density contrast distribution. Finally we obtain the corresponding linear values for matter used to build the theoretical curve: $\delta_\mathrm{v}=-0.887^{+0.038}_{-0.028}$. While there might be ways to improve the resizing procedure and reduce this uncertainty, we have focused here on testing this first version that relies on simple principles. We will explore possible improvements in future work. 

Within the estimated theoretical errors, we obtain a good agreement between data and theory without the need of free parameters, using only quantities that can be theoretically predicted: {\it we are able to forecast the void size function with a fully theoretical approach} \citep{svdw2004,jennings, contarini2019}. We note that, for low values of the resized radius, the measured void abundances are below theoretical predictions. This is a consequence to the sparse statistics of small voids when approaching the mean halo separation of the considered simulations; denser surveys will have access to better void statistics in those ranges (\eg PFS \citep{pfs2016}, WFIRST \citep{wfirst_2015}), likely filling the gap. 

Finally we check how the theoretical void size function depends on the threshold value. We vary the threshold value $\delta_\mathrm{v,NL}^H$ while holding the effective bias fixed (right panel Fig. \ref{Vdn_lcdm}). All the measured void abundances are within the uncertainty of the theoretical predictions (except at very small radii). This result confirms that the theoretical void size function prediction works for many threshold values. We notice that in observations this is a powerful tool to use, as it will allow to optimise the choice of the threshold to enhance cosmological constraints from the void size function.

The void size function prediction is obviously bias dependent, since the threshold $\delta_\mathrm{v}$ is derived from the ratio $\delta_\mathrm{v,NL}^\mathrm{H}/b_\mathrm{eff}$; therefore the use of different halo bias models can impact the theoretical prediction (we tested models such as \eg \citep{mo_e_white1996,warren_2004,efstathiou1988,cole_kaiser1988} instead of \citep{sheth_tormen1999}). The bias dependence of theoretical void size function might be affected by different systematic effects with respect to the halo mass function or observables related to clustering statistics. Interestingly, the use of the bias dependence of the void size function could be complementary to classical bias estimators. On the other hand, other techniques exist to directly estimate bias from data (\eg 3-pt halo statistics), and therefore it would be possible to rely on direct bias measurements rather than using theoretical bias models. This, of course, makes void size function predictions case-by-case dependent, since they are expected to change with the considered void tracers. The methodology, relying on bias measurements from higher order statistics, would be fully based on real data analysis. In future work we plan to directly measure the bias from our simulations to test this data-only based methodology with different cosmologies (see \cite{contarini2019} for the \lcdm case, albeit with a smaller simulation volume).

\subsubsection{Void abundance match: varying the dark energy EoS}

\begin{figure}[t]
\centering
\includegraphics[width=\textwidth]{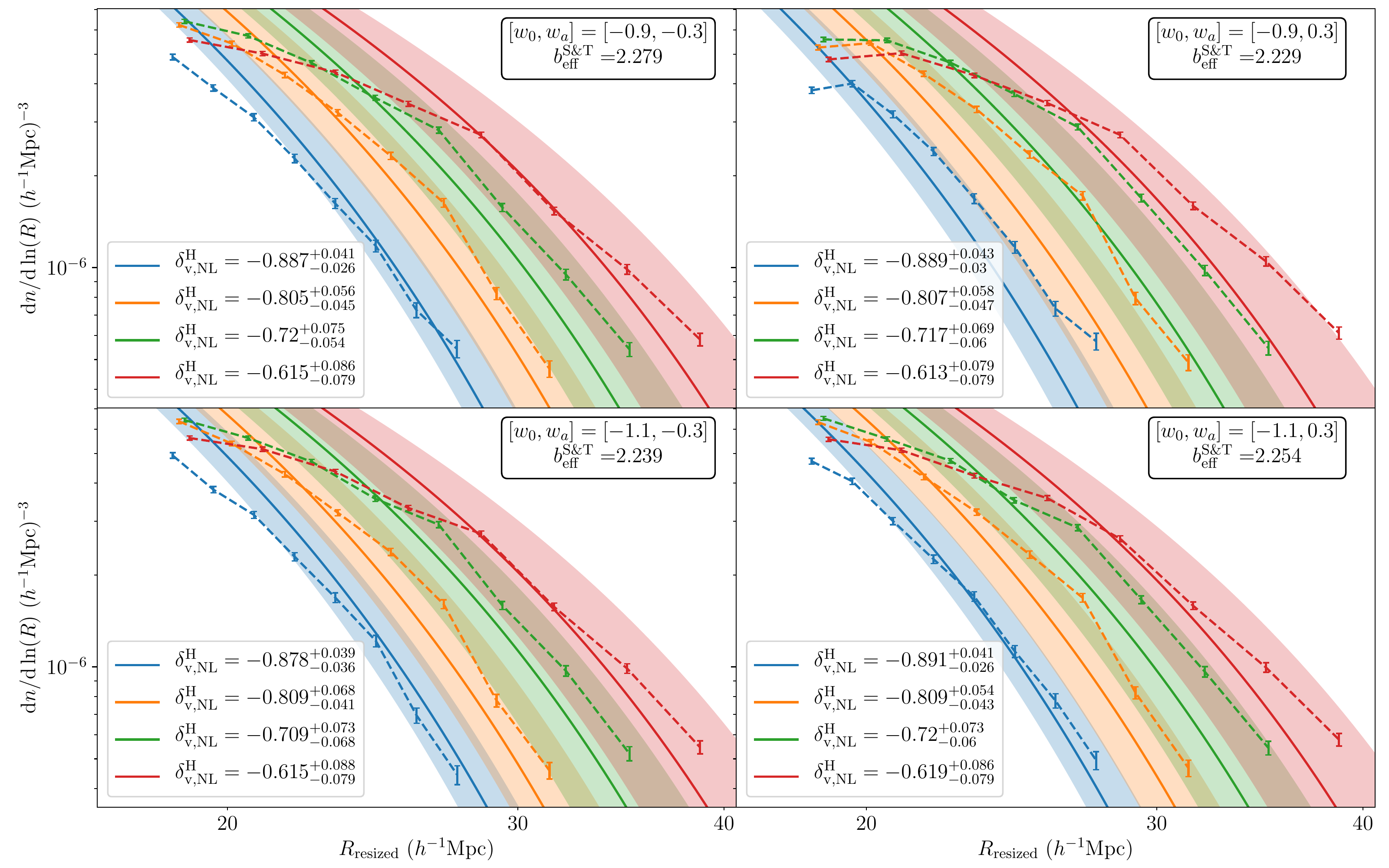}
\caption{Each panel represents the measured void abundances after applying the resizing procedure (dashed curves), and the corresponding theoretical void size functions (solid curves), for various threshold values $\delta_\mathrm{v,NL}^\mathrm{H}$, and for the four sets of parameters of the CPL dark energy EoS. Shaded areas give the uncertainty in the resizing procedure (see text).}
\label{Vdn_wowa}
\end{figure}

We repeat the above procedure for the CPL parametrisation with the four sets of $[w_0,w_a]$ parameters implemented in DEMNUni suite to check if the theoretical void size function is able to predict simulations in dynamical dark energy models. The results presented in Fig. \ref{Vdn_wowa} show we obtain a good agreement between the measurements and the theoretical void size functions in the different cosmologies, where the dependence in the void size function is implicitly contained in the linear growth factor $D(z)$ \citep{percival2005}, and mildly in the variance $\sigma$ of the linear matter perturbations. As for the \lcdm case, for all the analysed threshold values the agreement is obtained by using the Sheth \& Tormen \citep{sheth_tormen1999} effective bias, computed for all the different dark energy EoS. We note also in this case that the abundance of small voids falls below the theoretical prediction. 
In addition, mainly for the $[w_0,w_a]=[-0.9,0.3]$ case (upper right panel in Fig. \ref{Vdn_wowa}), the observed void abundance for two threshold values (green and orange curves) lies somewhat outside the uncertainty of the theoretical prediction. This mild mismatch could be due both to the non-optimal resizing procedure, and to bias modelling, the latter would need to be tested against different cosmological models.

Finally we notice that the resizing procedure reduces the dependence of the void abundance on the dark energy parametrisation. This is because this procedure finds the radius at which the mean number density of tracers within a sphere, centred on the void centre, reaches the considered threshold value. If this is not done precisely enough, one obtains the radius of voids corresponding to a distribution of thresholds, rather than one defined value (as the shaded area shows). In this way the differences due to different dark energy EoS can be washed out. However, the idea of void resizing is a promising procedure and deserves future investigation aiming at enhancing its precision. 
As discussed in Sec. \ref{conclu}, one simple strategy to reduce its uncertainty could be to rely on a more robust modelling of the void density profile (see \eg \cite{hamaus_density,massara_2018}). Further improving such precision is needed in view of forthcoming galaxy surveys such as DESI, Euclid, and WFIRST.

\bigskip

Summarising, we showed that the theoretical void size function agrees, within the errors, with measurements from simulations for different cosmological models, \ie the standard \lcdm model and the CPL parametrisation with the four sets of $[w_0,w_a]$ parameters implemented in DEMNUni. We also obtained a good agreement using various threshold values for each cosmological model. This allows on the one hand to select the best threshold $\delta_\mathrm{v}$ for the available data, and on the other hand to use different threshold values to better constrain cosmological parameters with void abundances (even if, of course, measurements in this case will be correlated).

Our work opens up the possibility to robustly predict void abundances from surveys in a fully theoretical way, even for various dark energy EoS, and to extract cosmological parameter constraints via void abundance from upcoming galaxy surveys.

\section{Conclusions and future perspectives}\label{conclu}
In this work we have confirmed, using halo catalogues from numerical simulations, the expectation that voids are sensitive to the dark energy EoS. Our analysis -- made possible by the large volume and resolution of DEMNUni simulations -- demonstrates that void abundance represents a robust diagnostic to test cosmology. As we have shown, the void size function can contribute to break degeneracies existing in other measurements between different dark energy EoS models, thus providing a complementary tool to classical cosmological probes.

We showed that the theoretical framework to describe the void size function is robust to be used with upcoming data-sets featuring a large volume and high statistics for large voids (\eg an Euclid-like survey). Our theoretical prediction is in good agreement with the actual simulation measurements, for all dark energy EoS considered and, notably, for different threshold values $\delta_\mathrm{v}$. 
Importantly, such an agreement is obtained using voids from halo catalogues, an approach that is closer to real galaxy catalogues and extends the work of Sheth \& van de Weygaert \citep{svdw2004}, van de Weygaert \& Bond \citep{vandeWeygaert-bond2008}, Jennings et al. \citep{jennings}, Pisani et al. \citep{pisani2015}, Ronconi et al. \citep{ronconi2019}, and Contarini et al. \citep{contarini2019}. Our theoretical approach is based on the halo bias model of Sheth \& Tormen \citep{sheth_tormen1999}; it could be interesting to explore the use of bias obtained directly from real data or simulations with different cosmologies. 
These results set the general framework to forecast with a fully theoretical approach the number density of voids as a function of their size to be expected from future surveys. They pave the way towards a reliable use of void abundance as a probe of cosmology. Future surveys such as DESI, Euclid, and WFIRST will observe a wider redshift range (and hence a larger volume) than the one considered in this work.
This will yield improved statistics and thus tighter constraints. 

Improvements in the methodology are still possible to further enhance its reliability. First, one could improve the treatment of Poisson voids, for example using a machine learning-based multivariate analysis as done in Cousinou et al. \citep{cousinou2018}).  These authors show that the density of the central Voronoi cell (reasonably correlated with the central density used here) and the effective radius  play a central role in discerning how reliable a void is, in particular for well-sampled catalogues as in our case (see \cite{cousinou2018} for details). A promising approach to reduce spurious voids that would deserve attention in upcoming work is the use of a first-order Delaunay Tessellation Field Estimator \citep{platen2007,vandeWeygaert2007}. For voids probed by a few halos only, the flat density profile in a Voronoi cell involves a noisy density field: taking the dual of the Voronoi tessellation to obtain a first-order density field would be more successful in interpolating the density field along ``empty'' void regions and would provide more reliable small voids in low-populated regions \citep{platen2007,vandeWeygaert2007}.
A second possibility, that we are considering for future work, would be to improve the modelling of density profiles (see \eg \citep{hamaus_density,massara_2018}, or, also,  \citep{cautun2016}, that defines the profile on the basis of void boundaries) to enhance the accuracy of the resizing procedure for each $\delta_\mathrm{v}$. This would allow for improved theoretical predictions and enhanced signal-to-noise, with a more powerful distinction between different dark energy EoS.

Investigations of the redshift dependence of the void size function in different dark energy scenarios and in combination with massive neutrinos, as well as the study of the impact of peculiar velocities \citep{pisani2015_rds}, are all possible using the extended set of DEMNUni simulations, and are left for future work. 
In this work we exploited the large volume of the DEMNUni simulations to study the abundance of voids and its use for cosmology, focusing on the void population captured by their volume and resolution. For future work it would be extremely interesting to study the low void mass/volume tail of the void abundance relying on simulations with large mass resolution. This would be particularly relevant for surveys deeply sampling the galaxy field, such as PFS, WFIRST or the Euclid deep fields. The dynamics of small voids being dominated by environment effects, further investigations would be required to test the use of the spherical approximation \citep{platen2008} in order to model the low volume/mass tail of the void abundance.
Finally further important aspects, such as the impact of survey mask and selection effects, will also deserve careful consideration in this context.

\acknowledgments
We thank the anonymous referee for helpful comments that allowed us to improve the quality of this manuscript. We thank Christina Kreisch for fruitful comments. 
GV acknowledges support from Universit\`a degli Studi di Padova and INFN sezione di Padova.
AP is supported  by  NASA  grant  15-WFIRST15-0008  to  the WFIRST Science Investigation Team ``Cosmology with the High Latitude  Survey".
CC and LG acknowledge support of the European Research Council through grant n. 291521 ``Darklight",  the Italian Space Agency (ASI grant I/023/12/0) and Italian MIUR (PRIN 2015  ``Cosmology and Fundamental Physics: Illuminating the Dark Universe with Euclid''.
NH acknowledges support from the DFG cluster of excellence ``Origins".
The DEMNUni simulations were carried out in the framework of the ``The Dark Energy and Massive-Neutrino Universe" project, using the Tier-0 IBM BG/Q Fermi machine and the Tier-0 Intel OmniPath Cluster Marconi-A1 of the Centro Interuniversitario del Nord-Est per il Calcolo Elettronico (CINECA). We acknowledge a generous CPU and storage allocation by the Italian Super-Computing Resource Allocation (ISCRA).

\bibliographystyle{JHEP.bst}
\bibliography{voids}

\end{document}